\documentclass{elsart}
\usepackage{graphicx}
\usepackage{amsmath}
\usepackage{amssymb}
\usepackage{dcolumn}
\usepackage{bm}

\def\vec#1{\mbox{\boldmath $#1$}}
\def\sp{/\!\!\!p}

\begin{document}

\begin{frontmatter}

\title{Exotic Quark Structure of $\Lambda$(1405) in QCD Sum Rule}

\author[address1]{T.Nakamura\thanksref{thank1}},
\author[address1]{J.Sugiyama\thanksref{thank2}},
\author[address2]{N.Ishii\thanksref{thank3}},
\author[address1]{T.Nishikawa\thanksref{thank4}}
and
\author[address1]{M.Oka\thanksref{thank5}} 

\address[address1]{Department of Physics, H-27, Tokyo Institute of Technology,\\
 Meguro, Tokyo 152-8551, Japan}

\address[address2]{Center for Computational Science, University of Tsukuba,\\
Tsukuba, Ibaraki 305-8571, Japan }

\thanks[thank1]{ 
E-mail: nakamura@th.phys.titech.ac.jp}
\thanks[thank2]{
sugiyama@th.phys.titech.ac.jp}
\thanks[thank3]{
ishii@ribf.riken.jp}
\thanks[thank4]{
nishi@th.phys.titech.ac.jp}
\thanks[thank5]{
oka@th.phys.titech.ac.jp}
\begin{abstract}
A new analysis is performed in QCD sum rule for the lightest negative
parity baryon $\Lambda (1405)$.
Mixings of three-quark and five-quark Fock components are taken into account.
Terms containing up to dimension 12 condensates are computed in the
operator product expansion.
It is found that the sum rule gives much stronger coupling of
$\Lambda^*$ to the five-quark operator
so that the five-quark components occupy about $90\%$ of
$\Lambda (1405)$.
\end{abstract}

\end{frontmatter}

The lightest negative parity baryon, $\Lambda^* (1405, I=0, J^{\pi}=1/2^-)$, 
is quite unique and mysterious in the baryon spectroscopy~\cite{Yao:2006px}.  
In the conventional quark model, 
it is supposed to consist of three quarks, $u-d-s$, and to have one unit of angular momentum 
so as to acquire negative parity. In the ground-state spectrum, 
the non-strange baryons are significantly lighter than the strange ones. 
This indicates that the constituent mass of the strange quark is 
about 150 MeV larger than that of the non-strange quarks.
It is not easy to explain why $\Lambda(1405)$ is significantly lighter than non-strange negative parity baryons, 
the lowest one of which is $N(1520)$.  
Assigning it to the flavor SU(3) singlet state is one solution to this problem, 
but the puzzle remains as it requires a large breaking of the flavor-spin SU(6), on the other hand.

Another puzzle is the size of the spin-orbit splitting for $\Lambda(1405)$.  
As its orbital angular momentum is one and the spin is (at least) 1/2, 
the spin-orbit partner state $J^{\pi}= 3/2^-$ is expected to exist.
A possible candidate is $\Lambda(1520)$. The splitting, 115 MeV, however,  
is  much larger than the spin-orbit splittings of the other members of the $p$-wave baryons, 
ex., $N(1535)-N(1520)$, or $\Sigma(1620)-\Sigma(1670)$.
The difficulty is also demonstrated in the quark models quantitatively~\cite{Reinders:1978mv,Reinders:1980af},
 namely, 
the standard (non-relativistic) quark model predicts the $p$-wave flavor singlet $\Lambda$ 
at about 50-150 MeV higher than the observed mass, and also a small spin-orbit splittings of order 10 MeV.

Under these circumstances, many conjectures on the structure of $\Lambda(1405)$ have been made. 
One possibility is that it is a bound state of  $\bar K$ and $N$ with the binding energy of about 28 MeV~\cite{Sakurai:1960ju,Dalitz:1967fp}.   
This possibility attracts a lot of attention recently in the context of possible strong $\bar K-N$ attraction, 
in particular inside nuclear medium~\cite{YA1,Akaishi:2002bg,Yamazaki:2002uh}. On the other hand, 
if the $S$-wave attraction is so strong that the $\bar K-N$ system is bound by 28 MeV, 
it is expected that their quark wave functions overlap significantly and the state looks more like a penta-quark.
Indeed, the 5-quark system may solve one of the above difficulties.  
Because the system of $4q-\bar q$ has negative intrinsic parity, 
they are all in the $L=0$ orbit and therefore no spin-orbit partner is necessary. 
Although it does not explain why the lowest negative parity baryon has
non-zero strangeness, we may argue that the corresponding non-strange
baryons can decay into $N\pi$ or $\Delta\pi \to N\pi\pi$ and are too
broad to be observed.

In this letter, we attempt to study possible 5-quark structure of $\Lambda^*$ directly from QCD.  
In the spectroscopy of hadrons, QCD is notorious to be non-perturbative and highly complicated due to the strong coupling as well as the confinement of color.
We here employ the QCD sum rule technique~\cite{Shifman:1978bx,Reinders:1984sr,Narison:1989aq,Narison:2002pw}
,which is applicable to non-perturbative regime of QCD to get information on the mass and structure of low-lying hadrons.
This method is considered as complimentary to the lattice QCD calculation, which is another popular approach for non-perturbative QCD. 
While the lattice QCD is limited currently by relatively large quark masses, the sum rule is capable to treat light (or massless) quarks.
Sum rule analyses employing 3-quark or 5-quark operators have been carried out in literatures
\cite{Liu:1984dp,Leinweber:1989hh,LSR3,Choe:1997wz,Jido:1996zw}. 
Most of them conclude that the mass of $\Lambda$(1405) is reproduced.
In the present study, we concentrate on the mixing of 3-quark and 5-quark Fock states in the sum rule.
In particular, we employ a flavor-singlet 5-quark operator made from scalar diquark operators, that is
considered to be preferred by the QCD interactions.

 In sum rule analyses, we calculate two point functions:
\begin{eqnarray}
\Pi(p)&=&i\int\! d^4\!x e^{ipx}\langle0|T[J(x)\bar{J}(0)]|0\rangle
\nonumber\\
&=&\Pi_1(p^2)+\Pi_p(p^2)\sp,
\label{s1}
\end{eqnarray}
where $J$ is an interpolating field operator that couples to the baryon state in question.
The functions $\Pi_1(p^2)$ and $\Pi_p(p^2)$ are expressed in two ways. In one side, they 
are evaluated by using the operator
product expansion~(OPE), which is applicable in deep Euclidean region, $-p^2\to\infty$, 
and are expressed in terms of the QCD parameters, such as the vacuum condensates, the current
quark masses and so on.  
On the other hand, the two point function is represented by a phenomenological
parametrization of the spectral function. The spectral function at the physical region($p^2>0$)
is assumed to have a sharp peak resonance for the ground state at $p^2=m^2$ and continuum 
contributions at $p^2>s_{\mathrm{th}}$. 
Using analyticity in the $p^2$ plane, they are related by the dispersion integral,
\begin{eqnarray}
\Pi^{\mathrm{OPE}}_{1}=\int^{\infty}_0 \!ds\frac{\rho^{\mathrm{OPE}}_{1}(s)}{s-p^2}
+(\textrm{subtraction terms}),
\nonumber \\
\Pi^{\mathrm{OPE}}_{p}=\int^{\infty}_0 \!ds\frac{\rho^{\mathrm{OPE}}_{p}(s)}{s-p^2}
+(\textrm{subtraction terms}),
\label{s2}
\end{eqnarray}
where $\rho^{\mathrm{OPE}}_{1(p)}=\textrm{Im}\Pi^{\mathrm{OPE}}_{1(p)}/\pi$
which are the spectral functions.
As usual, we parameterize the phenomenological spectral density as a single sharp pole
representing the negative parity resonance plus the continuum contributions given by 
the results obtained with the OPE, 
\begin{eqnarray}
\rho_1^{\mathrm{phen}}(s)&=&\frac{1}{\pi}\mathrm{Im}\;\Pi^{\mathrm{phen}}_1(s)=
-|\lambda|^2m\delta(s-m^2)+\Theta(s-s_{\mathrm{th}})\rho_1^{\mathrm{OPE}},
\nonumber \\
\rho_p^{\mathrm{phen}}(s)&=&\frac{1}{\pi}\mathrm{Im}\;\Pi^{\mathrm{phen}}_p(s)=
|\lambda|^2\delta(s-m^2)+\Theta(s-s_{\mathrm{th}})\rho_p^{\mathrm{OPE}},
\label{s3}
\end{eqnarray} 
where $\lambda$ is coupling strength of the physical state under investigation. 
The minus sign in the first term of $\rho_1^{\mathrm{phen}}$ is a signature for negative 
parity of $\Lambda$(1405). The sum rule is obtained by matching two expressions, so that
the mass of the resonance, $m$, and the other phenomenological parameters can be determined
from the QCD parameters. In order to extract the ground state information by suppressing 
the continuum contributions, we apply the Borel transformation,
\begin{eqnarray}
\mathcal{B}\equiv \lim_{-p^2,n\to \infty}\frac{(-p^2)^{n+1}}{n!}\left(\frac{d}{dp^2}\right)^n,
\label{s4}
\end{eqnarray}
where the limit is taken with the Borel mass $M_B^2\equiv-p^2/n$ fixed. Due to the $p^2$ derivative,
the subtraction terms, i.e. a finite polynomial in $p^2$, in Eq. (\ref{s2}) vanish. 
The Borel transformation also suppresses the effects of the 
higher dimensional terms in the OPE. Then we arrive at the sum rule equations,
\begin{eqnarray}
|\lambda|^2me^{-m^2/M_B^2}&=&-\int_0^{s_{\mathrm{th}}}\!dse^{-s/M_B^2}\rho_1^{\mathrm{OPE}}(s),
\nonumber \\
|\lambda|^2e^{-m^2/M_B^2}&=&\int_0^{s_{\mathrm{th}}}\!dse^{-s/M_B^2}\rho_p^{\mathrm{OPE}}(s).
\label{s5}
\end{eqnarray}
Taking the ratio of the resulting equation and its first derivative with respect to $1/M_B^2$,
we obtain the ground
state mass $m$ as function of the two parameters $s_{\mathrm{th}}$ and $M_B$. Ideally, the ground
state mass should have weak dependence of the two parameters. In order to obtain the reliable 
conclusion, the dependences should be checked carefully.

We employ the following interpolating fields for the flavor singlet $\Lambda$(1405),
\begin{eqnarray}
J_3&=&\epsilon_{abc}\Biggl[\left(u_a^TC\gamma_5d_b\right)s_c-
\left(u_a^TCd_b\right)\gamma_5 s_c
-\left(u_a^TC\gamma_5\gamma^{\mu}d_b\right)\gamma_\mu s_c\Biggl]
\nonumber \\
J_5 &=& \epsilon_{abc}\epsilon_{def}\epsilon_{cfg}\Biggl[\left(d_a^TC\gamma_5s_b\right)
\left(s_d^TC\gamma_5u_e\right)\gamma_5C\bar{s}_g^T
\nonumber \\
&+&\left(s_a^TC\gamma_5u_b\right)
\left(u_d^TC\gamma_5d_e\right)\gamma_5C\bar{u}_g^T
+\left(u_a^TC\gamma_5d_b\right)
\left(d_d^TC\gamma_5s_e\right)\gamma_5C\bar{d}_g^T
\Biggr],
\label{s6}
\end{eqnarray}
where $a,b,\cdots$ represent colors and $C=i\gamma^2\gamma^0$. The 3-quark interpolating field 
is uniquely determined. 
On the other hand, the 5-quark interpolating field has variations. This form is taken because
we expect the scalar diquarks favored in the multi-quark components.

Our results of OPE are listed in Appendix \ref{ap.A}. The OPE is calculated up to
the dimension-6 terms for $\Pi_{33}(p)$. 
In order to make the sum rule consistent in the power expansion
in $1/p^2$, the correlators of higher dimensional operators require
higher dimensional terms.
Thus, the OPE of $\Pi_{35}(p)$ is calculated up to the terms with
dim.-9 condensates and $\Pi_{55}(p)$ up to dim.-12 terms.
As usual, we assume the vacuum saturation approximation for the higher dimensional operators
such as $\langle\bar{q}\bar{q}qq\rangle$ and $\langle\bar{q}\bar{q}qqG\rangle$.

We consider the diagrams containing the $q\bar{q}$ annihilation because it is necessary to
deal with the mixing of different Fock state. 
The systematic calculation is possible in the framework of the sum rule, {\it{i.e.}},
the quark pair annihilations are substituted for $\langle\bar{q}q\rangle$ or 
$\langle\bar{q}g_s\vec{\sigma}\!\cdot\!\vec{G}q\rangle$. Because the interpolating fields are 
under the normal ordering, the perturbative part of the $q\bar{q}$ annihilation must disappear.
Because the OPE is represented as a polynomial in x, only the zero-th order term survives in
the $x\to 0$ limit.

The main purpose of this study is to investigate whether $\Lambda(1405)$ is dominated 
by $qqq$ or $qqqq\bar{q}$, and, if they are mixed, to calculate their probabilities.
It would be natural to consider the strengths of the couplings of the 3-quark and 5-quark
operators to the physical state and then evaluate the mixing angle. However , such a procedure is 
largely dependent on the definition and normalization of the local operators. We have proposed
two ways to ``define" the ratio of the Fock space probability in analysis of the scalar mesons
\cite{Sugiyama:2007sg}. The definition of mixing is applied to $\Lambda(1405)$ analysis.

In the first approach, we define local operators tentatively ``normalized" in the context of a
full 5-quark operator $J_5$ in Eq.(\ref{s6}). In fact, $J_5$ contains a part effectively reduced
to $J_3$ multiplied by $\bar{q}q$. Thus we define the $J_3'$ and $J_5'$ by
\begin{eqnarray}
J_5=J_5'
+\underbrace{\left(-\frac{1}{18}(\langle \bar{u}u\rangle
                +\langle \bar{d}d\rangle
                +\langle \bar{s}s\rangle)J_3\right)}_{\Large J^\prime_3}.
\label{s7}
\end{eqnarray}
We regard $J_3'$ and $J_5'$ as  ``normalized" 3-quark and 5-quark operators, respectively.
Using $J_3'$ and $J_5'$, we define the mixing parameter, $\theta_1$, by
\begin{eqnarray}
\langle 0|J_3'(x)|\Lambda^\ast\rangle&=&\lambda\cos\theta_1\gamma_5u(x),
\nonumber \\
\langle 0|J_5'(x)|\Lambda^\ast\rangle&=&\lambda\sin\theta_1\gamma_5u(x),
\label{s8}
\end{eqnarray}
where $\lambda$ is the coupling strength and $u(x)$ is the Dirac spinor for $\Lambda^*$ state.
This definition happens to be equal to defining $\theta_1$ so that 
$J_\Lambda(x)=\cos\theta J_3'(x)+\sin\theta J_5'(x)$ couples to the physical state most strongly
at $\theta=\theta_1$, i.e.,
\begin{eqnarray}
\langle 0|J_{\Lambda}(x)|\Lambda^\ast\rangle
=\lambda (\cos\theta\cos\theta_1+\sin\theta\sin\theta_1) \gamma_5u(x) \nonumber\\
=\lambda \cos(\theta-\theta_1) \gamma_5u(x)
\stackrel{\theta=\theta_1}{\Longrightarrow} \lambda \gamma_5u(x).
\label{s9}
\end{eqnarray}

The mixing 
parameter $\theta_1$ can be evaluated from the correlation functions under the assumption
that the correlators for $J_3'$ and $J_5'$ show the resonance pole at the same position. 
In the case of the chiral odd correlator $\Pi_1(q^2)$, we obtain
\begin{eqnarray}
\frac{1}{\pi}\textrm{Im}\Pi_1^{33'}(q^2)=-m|\lambda_{33}|^2\delta(q^2-m^2)+\textrm{cont.},
\nonumber \\
\frac{1}{\pi}\textrm{Im}\Pi_1^{55'}(q^2)=-m|\lambda_{55}|^2\delta(q^2-m^2)+\textrm{cont.},
\label{s10}
\end{eqnarray}
where $\Pi^{33'}(q^2)$ and $\Pi^{55'}(q^2)$ are the correlation functions by using the operators $J_3'$ and $J_5'$.
Using
\begin{eqnarray}
|\tan\theta_1| = \frac{|\lambda_{55}|}{|\lambda_{33}|},
\label{s10-1}
\end{eqnarray}
we can extract the resonance pole position $m$ and the mixing parameter
$\theta_1$
simultaneously.
We can also calculate the mixing parameter $\theta_1$ from the chiral
even correlators.
In the actual calculation of $\lambda_{33}$ and $\lambda_{55}$,
we apply the Borel transform on Eq.(\ref{s10}), and use the sum rule
relations as in Eq.({\ref{s5}).

This definition of the mixings is model independent, but
it depends on the choice of the local operators, which define the
normalization of the
$J_3'$ and $J_5'$.
Therefore it does not necessarily have a direct
relation to the mixing parameter employed in the quark models.
In order to define mixing parameter in a way more directly connected to
the quark models, one should normalize
the operators using quark model wave functions. For instance, the local
operators, Eq.(\ref{s6}),
can be normalized to the wave functions of 3-quark and 5-quark states in
the MIT bag model~\cite{DeGrand:1976kx,Strottman:1979qu},
\begin{eqnarray}
|\Lambda_{3q}\rangle
&=&|SSP\rangle
\otimes\frac{1}{\sqrt{2}}\left(|\uparrow\downarrow\uparrow\rangle-|\downarrow\uparrow\uparrow\rangle\right)
\otimes
\frac{1}{\sqrt{6}}\epsilon_{\alpha\beta\gamma}|\alpha\beta\gamma\rangle
\otimes\frac{1}{\sqrt{6}}\epsilon_{abc}|abc\rangle,
\nonumber \\
|\Lambda_{5q}\rangle
&=&|SSSSS\rangle\otimes\frac{1}{2}\left(
|\uparrow\downarrow\uparrow\downarrow\uparrow\rangle
-|\uparrow\downarrow\downarrow\uparrow\uparrow\rangle
-|\downarrow\uparrow\uparrow\downarrow\uparrow\rangle
+|\downarrow\uparrow\downarrow\uparrow\uparrow\rangle
\right)
\nonumber\\
&&
\otimes\frac{1}{2\sqrt{6}}\epsilon_{\alpha\beta\gamma}\epsilon_{\delta\rho\omega}\epsilon_{\gamma\omega\lambda}|\alpha\beta\delta\rho\bar{\lambda}\rangle
\otimes\frac{1}{2\sqrt{6}}\epsilon_{abc}\epsilon_{def}\epsilon_{cfg}|abdeg\rangle,
\label{s11}
\end{eqnarray}
where $\alpha,\beta,\cdots$ represent the flavors and $a,b,\cdots$
represent the colors.

As the bag model states (with definite number of quarks) are properly
normalized,
the matrix elements,
\begin{eqnarray}
\langle 0|J_3(0)|\Lambda_{3q}\rangle =\lambda_3 \gamma_5u(0),
\nonumber\\
\langle 0|J_5(0)|\Lambda_{5q}\rangle=\lambda_5 \gamma_5u(0),
\label{s12}
\end{eqnarray}
gives ``normalized'' operators, $J_3/\lambda_3$ and $J_5/\lambda_5$.
The rest of the procedure is identical to the one give above for $J_3'$
and $J_5'$, in
Eqs.(\ref{s8}-\ref{s10-1}).

In calculating the mixing parameter,
we needs only the ratio of $\lambda_3$ and $\lambda_5$, which is given by
\begin{eqnarray}
\frac{\lambda_5}{\lambda_3}&=&-\frac{\sqrt{2}}{8\pi}\frac{{\mathcal{N}_5(S_{1/2})}^5}
{{\mathcal{N}_3(S_{1/2})}^2\mathcal{N}_3(P_{1/2})}\sim
-0.17\frac{R_3^{9/2}}{R_5^{15/2}},
\label{s13}\\
\mathcal{N}_n(S_{1/2})&=&\frac{ER_n}{R_n^{3/2}|j_0(ER_n)|}\frac{1}{\sqrt{2ER_n(-1+ER_n)}},
\nonumber \\
\mathcal{N}_n(P_{1/2})&=&\frac{ER_n}{R_n^{3/2}|j_1(ER_n)|}\frac{1}{\sqrt{2ER_n(1+ER_n)}},
\nonumber \\
&&\textrm{and}\hspace{2em}ER_n=
\left\{ \begin{array}{ll}
2.04 &  \hspace{2em}\textrm{for}\hspace{1em}S_{1/2}\\
3.81 &  \hspace{2em}\textrm{for}\hspace{1em}P_{1/2}\\
\end{array} \right.,
\nonumber
\end{eqnarray}
where $R_n$ is the bag radius of the $n$-quark states and gives the
dimensional scale of the normalizations.
The ratio $\lambda_5/\lambda_3$ depends on the bag radii because the
operators $J_3$ and $J_5$ have
different dimensions.
We here assume that the bag radius of the p-wave 3-quark state
is same as that of the s-wave 5-quark state.

It is well-known that the baryon correlator contains contributions from both positive and
negative parity baryons. It is possible to project out the parity and study the spectrum of
positive and negative parity state separately~\cite{Jido:1996ia}.
It is shown that neither the three-quark nor five-quark operators
give positive spectral functions for the positive parity $\Lambda^*$
states in the low energy region.
This implies that negative parity contribution is dominant for $\Lambda^*$
state in the low energy region. Thus we conclude that the lowest energy 
flavor singlet $\Lambda^*$ is a negative parity state.
This is consistent to experimental results. 
We show the masses of $\Lambda^*$ in the case of pure 3-quark and pure 5-quark
in FIG.\ref{Fig1}.
The values of QCD parameters are taken as $m_u=m_d=0$,
$m_s=0.12\rm{GeV}$, $\langle\bar{q}q\rangle
=(-0.23\rm{GeV})^3$, $\langle\bar{s}s\rangle=0.8\times\langle\bar{q}q\rangle$, 
$\langle\bar{q}g_s\sigma\!\cdot\!Gq\rangle/\langle\bar{q}q\rangle=0.8\rm{GeV}^2$
and $\langle\alpha_s\pi^{-1}G^2\rangle=(0.33\rm{GeV})^4$.
\begin{figure}[tb]
\begin{center}
\includegraphics[width=18pc]{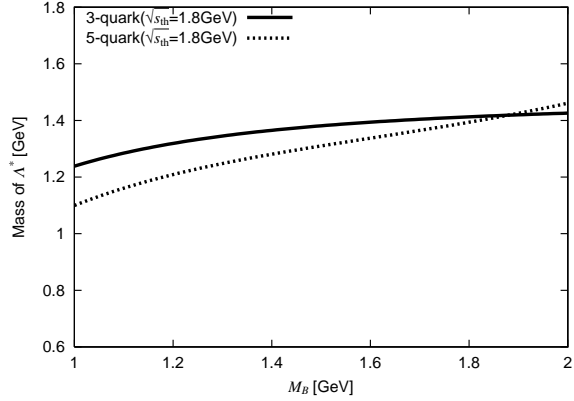}
\end{center}
\caption{The mass of $\Lambda^*$ in the case of pure 3-quark and pure 5-quark 
plotted as a function of the Borel mass, $M_B$. The threshold parameter, $\sqrt{s_{\rm{th}}}$,
is fixed to $1.8\rm{GeV}$. 
}
\label{Fig1}
\end{figure}
In this figure, we only present the sum rule from the $\Pi_1$ 
structure because the $\Pi_1$ sum rule is more reliable than the $\Pi_q$ sum rule.
We see that the Borel stability is fairly good. 
The pole positions of the 
3-quark and the 5-quark state are close to each other.
Therefore, we estimate the mass of the mixed $\Lambda^*$ defined 
by the first normalization method in the case of various threshold parameters, $s_{\rm{th}
}$, in FIG.\ref{Fig2}. 

\begin{figure}[tb]
\begin{center}
\includegraphics[width=18pc]{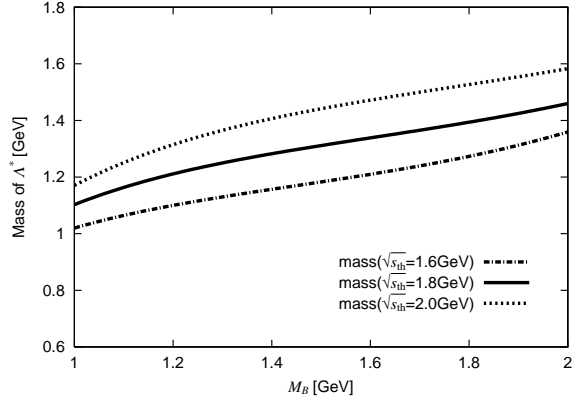}
\end{center}
\caption{The mass of $\Lambda^*$, where the 3-quark and 5-quark components are
mixed, plotted as a function of the Borel mass, $M_B$, for various threshold parameters,
$\sqrt{s_{\rm{th}}}=1.6$, $1.8$ and $2.0\rm{GeV}$.  
}
\label{Fig2}
\end{figure}

\begin{table*}[b]
\begin{tabular}{cccc}
\hline\hline
$\sin^2\theta_1$&$\sin^2\theta_2$($R_{\rm{bag}}=5.1\rm{GeV}^{-1}$) 
&$\sin^2\theta_2$($R_{\rm{bag}}=5.5\rm{GeV}^{-1}$)
&$\sin^2\theta_2$($R_{\rm{bag}}=5.9\rm{GeV}^{-1}$)\\
\hline
0.91 & 0.96 & 0.97 & 0.98\\
\hline\hline
\end{tabular}
\caption{\label{tab:table1}The mixing parameters from the first normalization method and the
normalized operators according to the bag model wave functions for various bag radii.
}
\end{table*}
The mass calculated from the mixed operator, $J_{\Lambda}$, is about $1.1\sim1.4\rm{GeV}$,
which is similar to the result from the pure 5-quark operator given in FIG.\ref{Fig1}.
In fact, we find that the mixed state is dominated by the 5-quark component. In TABLE 
\ref{tab:table1}, We show that 5-quark component calculated from the first normalization
method and the normalized operators according to the bag model wave functions for various bag 
radii. In the bag model, we take the central value as $R=5.5\rm{GeV}^{-1}$, which is the bag radius
determined for the 3-quark component\cite{DeGrand:1976kx}. 
We observe that the mixing parameters are almost independent of the
Borel mass, $M_B$.
We find that it is also independent of the threshold parameter, $s_{th}$.
The weak $M_B$-dependence of the mixing parameter may be attributed to the
cancellation of the $p$-dependences ($M_B$-dependences) of the 3- and
5-quark correlators,
while the mass extracted from the sum rule depends both on the threshold
parameter and
the Borel mass.
Thus we find that the 5-quark component occupies 90$\%$ of $\Lambda^*$
in the first normalization,
while the 5-quark component calculated by the bag model normalization
is also more than 90$\%$.

In order to check how well these sum rules work, we calculate the pole contribution
defined by
\begin{equation}
 \frac{\mathcal{B}\left[\Pi_{\mathrm{OPE}}(p^2)\theta(s_{th}-p^2)\right]}
{\mathcal{B}\left[\Pi_{\mathrm{OPE}}(p^2)\right]}. \label{s16}
\end{equation}
The pole contribution for the pure 5-quark correlator is less than the one for the pure 3-quark
correlator. The reason is that the OPE of the 5-quark correlator contains higher powers of $p^2$
and grows rapidly for large $p^2$. The pole contribution for mixed operator is almost the same
as the one for pure 5-quark operator.  
We take the valid Borel window as the region where the pole contribution is more than 30$\%$.
This constraint requires the Borel window for the mixing state, $M<1.4~\rm{GeV}$.
In FIG.\ref{Fig2}, it is seen that the mass of $\Lambda^*$ for the mixed operator is lighter than
the one of experimental results. We may adjust the threshold parameter to match it to the experimental
results. However, the mixing parameter is independent of the threshold parameter. Therefore the conclusion
that the 5-quark component is dominant in $\Lambda^*$ will not change.

We summarize the results in this work.
A formulation is proposed to take into account mixing of different
Fock states in the QCD sum rule. In order to quantify the mixing probability,
one needs the normalization of the operators. We suggest two ways: One is to define the 
normalization using a multiquark operator which couples to the ground state most strongly.
The other way is to adjust the operators to normalized wave functions from the MIT bag model.
We apply the formulation to the flavor singlet $\Lambda^*$. Our results indicate that $90\%$
or more of $\Lambda^*$ is composed of the 5-quark components.

Several groups performed quenched lattice QCD calculations for
$\Lambda(1405)$
~\cite{Nemoto:2003ft,Melnitchouk:2002eg,Lee:2002gn,Burch:2006cc,Ishii:2007ym}.
However, their results are controversial.
By using a flavor singlet operator, they have obtained qualitatively a
consistent result in the heavy  quark mass region $m_{\pi}^2 \leq 0.4$
GeV$^2$, i.e., their data in this region extrapolates to a mass, which
is significantly heavier than $\Lambda(1405)$,
suggesting  the  importance of  the  chiral  quark  effect and/or  the
coupling to the intermediate $N\bar{K}$ state.
Ref.~\cite{Lee:2002gn} and Ref.~\cite{Burch:2006cc} included a quenched lattice QCD calculation in the
light quark  mass region $m_{\pi}^2  \leq 0.2$ GeV$^2$ and  obtained a
low-lying $\Lambda(1405)$.
However, their results do not seem to be consistent with each other.
Ref.~\cite{Lee:2002gn} obtained  low-lying $\Lambda(1405)$ by using  a flavor singlet
operator with  the overlap fermion,  suggesting the importance  of the
chiral  quark   effect.   In  contrast, Ref.~\cite{Burch:2006cc} obtained  low-lying
$\Lambda(1405)$  with a  flavor octet  operator, whereas  their result
remains heavy with a flavor singlet operator, which indicates that the
coupling  to $N\bar{K}$  state plays  an important  role in  the light
quark mass region.
Ref.~\cite{Ishii:2007ym} made a  special analysis by performing a  quenched lattice QCD
calculation with a five quark operator of $N\bar{K}$ type omitting the
annihilation diagram.
Their data in the heavy quark mass region $m_{\pi}^2 \leq 0.4$ GeV$^2$
extrapolates to a state, which  is even higher than the conventional three
quark result.
Obviously, we  need more  information on $\Lambda(1405)$  to elucidate
its intrinsic nature.  It is  necessary to perform further lattice QCD
analyses using,  for instance, unquenched QCD,  light quark, couplings
among varieties of interpolating fields corresponding to different Fock 
components. 

It is interesting to compare our conclusion with recent
activities on the properties of $\Lambda(1405)$ in nuclear medium~\cite{YA1,Akaishi:2002bg,Yamazaki:2002uh}.
It has been argued that the strong attraction between $\bar K$ and $N$ 
makes a bound state, which is nothing but the $\Lambda$ resonance.
In that picture, the strong $S$-wave attraction makes the $N\bar K$ 
bound system quite compact according to the recent calculation.
We conjecture that such a picture of $\Lambda(1405) $ is 
consistent with our sum rule result.  Namely, the compact $N\bar K$
bound state looks just like a 5-quark state in QCD.  

Finally, for further analyses, it will be important to see how the
results depend on the choice of the five quark operator.
The present calculation employs a flavor-singlet operator which
is composed of two flavor $\bar 3$ diquarks and an antiquark.
The advantage of this operator is that it fits well with quark 
model picture of the isolated negative-parity 5-quark baryon.
From the QCD sum rule viewpoint, it is not necessarily
the most suitable operator for this state.  Further study 
along this line will be worthwhile.
    
{\it Acknowledgements.}

This work was supported in part by KAKENHI (17070002 and 19540275)
and the 21st Century COE Program at TokyoTech ``Nanometer-Scale 
Quantum Physics'' by the Ministry of Education, Culture, Sports, Science and Technology.

\appendix
\section{The Result of OPE}\label{ap.A}
The results of the OPE are summarized as:
\begin{eqnarray}
\Pi_{33}(p)&=&i\int\!\textrm{d}^4\!x\>e^{ipx}\langle 0|T[J_3(x)\bar{J}_3(0)]|0\rangle
\nonumber \\
&=&-\frac{1}{2^8\pi^4}p^4\sp\ln{(-p^2)}+\frac{m_u}{2^6\pi^4}p^4\ln{(-p^2)}
-\frac{\langle\bar{u}u\rangle}{2^3\pi^2}p^2\ln{(-p^2)}
\nonumber\\
&&-\frac{1}{2^4\pi^2}\Big(3m_u\langle\bar{u}u\rangle+2m_u\langle\bar{d}d\rangle
+2m_u\langle\bar{s}s\rangle\Big)\sp\ln{(-p^2)}
\nonumber\\
&&-\frac{1}{2^7\pi^2}\langle
\alpha_s\pi^{-1}G^2\rangle\sp\ln{(-p^2)}+\frac{\sp}{3p^2}\langle\bar{u}u\rangle
\langle\bar{d}d\rangle
\nonumber\\
&&+\frac{m_u\sp}{2^5\pi^2p^2}\Big(\langle\bar{d}g_s\vec{\sigma}\!\cdot\!\vec{G}d\rangle
+\langle\bar{s}g_s\vec{\sigma}\!\cdot\!\vec{G}s\rangle\Big)
\nonumber \\
&&\hspace{13em}+\{\textrm{flavor cyclic rotation}\},
\label{aeq1}
\end{eqnarray}
\begin{eqnarray}
\Pi_{35}(p)&=&i\int\!\textrm{d}^4\!x\>e^{ipx}\langle 0|T[J_3(x)\bar{J}_5(0)]|0\rangle
\nonumber \\
&=&\frac{\langle\bar{u}u\rangle}{3\!\times\!2^9\pi^4}p^4\sp\ln{(-p^2)}
-\frac{m_u\langle\bar{u}u\rangle}{3\!\times\!2^7\pi^4}p^4\ln{(-p^2)}
\nonumber\\
&&+\frac{\langle\bar{u}g_s\vec{\sigma}\!\cdot\!\vec{G}u\rangle}{3\!\times\!2^9\pi^4}
p^2\sp\ln{(-p^2)}
+\frac{\langle\bar{u}u\rangle^2}{3\!\times\!2^4\pi^2}p^2\ln{(-p^2)}
\nonumber\\
&&-\frac{m_u}{2^{10}\pi^4}\Big(2\langle\bar{u}g_s\vec{\sigma}\!\cdot\!\vec{G}u\rangle
+\langle\bar{d}g_s\vec{\sigma}\!\cdot\!\vec{G}d\rangle
+\langle\bar{s}g_s\vec{\sigma}\!\cdot\!\vec{G}s\rangle\Big)p^2\ln{(-p^2)}
\nonumber\\
&&+\frac{m_u\langle\bar{d}d\rangle\langle\bar{s}s\rangle}{3\!\times\!2^3\pi^2}\sp\ln{(-p^2)}
+\frac{\langle\alpha_s\pi^{-1}G^2\rangle}{3\!\times\!2^8\pi^2}\langle\bar{u}u\rangle\sp\ln{(-p^2)}
\nonumber\\
&&-\frac{1}{3\!\times\!2^5\pi^2}\langle\bar{u}u\rangle\langle\bar{u}g_s\vec{\sigma}\!\cdot\!
\vec{G}u\rangle\ln{(-p^2)}-\frac{\sp}{2p^2}\langle\bar{u}u\rangle\langle\bar{d}d\rangle
\langle\bar{s}s\rangle
\nonumber\\
&&+\frac{m_u\sp}{3^2\!\times\!2^8\pi^2p^2}\Big(10\langle\bar{u}u\rangle
\langle\bar{u}g_s\vec{\sigma}\!\cdot\!\vec{G}u\rangle-8\langle\bar{s}s\rangle
\langle\bar{u}g_s\vec{\sigma}\!\cdot\!\vec{G}u\rangle-8\langle\bar{d}d\rangle
\langle\bar{u}g_s\vec{\sigma}\!\cdot\!\vec{G}u\rangle
\nonumber\\
&&\hspace{5em}+3\langle\bar{u}u\rangle\langle\bar{d}g_s\vec{\sigma}\!\cdot\!\vec{G}d\rangle\Big)
+3\langle\bar{u}u\rangle\langle\bar{s}g_s\vec{\sigma}\!\cdot\!\vec{G}s\rangle
-18\langle\bar{d}d\rangle\langle\bar{s}g_s\vec{\sigma}\!\cdot\!\vec{G}s\rangle
\nonumber\\
&&\hspace{5em}-18\langle\bar{s}s\rangle\langle\bar{d}g_s\vec{\sigma}\!\cdot\!\vec{G}d\rangle
\Big)+\{\textrm{flavor cyclic rotation}\},
\label{aeq2}
\end{eqnarray}
\begin{eqnarray}
\Pi_{55}(p)&=&i\int\!\textrm{d}^4\!x\>e^{ipx}\langle 0|T[J_5(x)\bar{J}_5(0)]|0\rangle
\nonumber \\
&=&-\frac{3}{5!\;7!\;2^9\pi^8}p^{10}\sp\ln{(-p^2)}+\frac{3m_u}{5!\;6!\;2^9\pi^8}p^{10}
\ln{(-p^2)}-\frac{\langle\bar{u}u\rangle}{4!\;5!\;2^7\pi^6}p^8\ln{(-p^2)}
\nonumber\\
&&-\frac{m_u}{5\!\times\!2^{12}\!\times\!3^2\pi^6}\Big(5\langle\bar{u}u\rangle
-4\langle\bar{d}d\rangle-4\langle\bar{s}s\rangle\Big)p^6\sp\ln{(-p^2)}
\nonumber\\
&&-\frac{\langle\alpha_s\pi^{-1}G^2\rangle}{4!\;5!\;2^8\pi^6}p^6\sp\ln{(-p^2)}
+\frac{\langle\bar{u}g_s\vec{\sigma}\!\cdot\!\vec{G}u\rangle}{3^2\!\times\!2^{13}\pi^6}p^6\ln{(-p^2)}
\nonumber\\
&&-\frac{1}{3^2\!\times\!2^{12}\pi^4}\Big(3\langle\bar{u}u\rangle^2+17\langle\bar{u}u\rangle
\langle\bar{s}s\rangle\Big)p^4\sp\ln{(-p^2)}
\nonumber\\
&&-\frac{m_u}{3^2\!\times\!2^{16}\pi^6}\Big(8\langle\bar{u}g_s\vec{\sigma}\!\cdot\!
\vec{G}u\rangle+9\langle\bar{d}g_s\vec{\sigma}\!\cdot\!vec{G}d\rangle
+9\langle\bar{s}g_s\vec{\sigma}\!\cdot\!\vec{G}s\rangle\Big)p^4\sp\ln{(-p^2)}
\nonumber\\
&&-\frac{m_u}{3^2\!\times\!2^{10}\pi^4}\Big(-7\langle\bar{d}d\rangle^2-7\langle
\bar{s}s\rangle^2-9\langle\bar{u}u\rangle^2-23\langle\bar{u}u\rangle\langle\bar{s}s\rangle
-23\langle\bar{u}u\rangle\langle\bar{d}d\rangle
\nonumber\\
&&\hspace{5em}+\langle\bar{d}d\rangle\langle\bar{s}s\rangle\Big)p^4\ln{(-p^2)}
-\frac{\langle\bar{u}u\rangle}{3^2\!\times\!2^{10}\pi^4}\langle\alpha_s\pi^{-1}G^2\rangle
p^4\ln{(-p^2)}
\nonumber\\
&&-\frac{\langle\bar{u}u\rangle}{3^2\!\times\!2^{12}\pi^4}\Big(6\langle\bar{u}g_s
\vec{\sigma}\!\cdot\!\vec{G}u\rangle-31\langle\bar{d}g_s\vec{\sigma}\!\cdot\!\vec{G}d\rangle
-31\langle\bar{s}g_s\vec{\sigma}\!\cdot\!\vec{G}s\rangle\Big)p^2\sp\ln{(-p^2)}
\nonumber\\
&&-\frac{1}{3^2\!\times\!2^7\pi^2}\Big(\langle\bar{u}u\rangle^3+10\langle\bar{u}u\rangle
\langle\bar{s}s\rangle^2+17\langle\bar{u}u\rangle^2\langle\bar{s}s\rangle-
\langle\bar{u}u\rangle\langle\bar{d}d\rangle\langle\bar{s}s\rangle\Big)p^2\ln{(-p^2)}
\nonumber\\
&&+\frac{m_u}{3^2\!\times\!2^{13}\pi^4}\Big(88\langle\bar{u}u\rangle
\langle\bar{u}g_s\vec{\sigma}\!\cdot\!\vec{G}u\rangle
-105\langle\bar{u}u\rangle\langle\bar{s}g_s\vec{\sigma}\!\cdot\!\vec{G}s\rangle
-105\langle\bar{u}u\rangle\langle\bar{d}g_s\vec{\sigma}\!\cdot\!\vec{G}d\rangle
\nonumber\\
&&\hspace{2em}-214\langle\bar{s}s\rangle\langle\bar{u}g_s\vec{\sigma}\!\cdot\!\vec{G}u\rangle
-214\langle\bar{d}d\rangle\langle\bar{u}g_s\vec{\sigma}\!\cdot\!\vec{G}u\rangle
-201\langle\bar{s}s\rangle\langle\bar{s}g_s\vec{\sigma}\!\cdot\!\vec{G}s\rangle
\nonumber\\
&&\hspace{2em}-201\langle\bar{d}d\rangle\langle\bar{d}g_s\vec{\sigma}\!\cdot\!\vec{G}d\rangle
+12\langle\bar{d}d\rangle\langle\bar{s}g_s\vec{\sigma}\!\cdot\!\vec{G}s\rangle
+12\langle\bar{s}s\rangle\langle\bar{d}g_s\vec{\sigma}\!\cdot\!\vec{G}d\rangle
\Big)p^2\ln{(-p^2)}
\nonumber\\
&&-\frac{m_u}{3^2\!\times\!2^8\pi^2}\Big(3\langle\bar{u}u\rangle^3-2\langle\bar{d}d\rangle^3
-2\langle\bar{s}s\rangle^3-3\langle\bar{u}u\rangle\langle\bar{d}d\rangle\langle\bar{s}s\rangle
+15\langle\bar{d}d\rangle\langle\bar{u}u\rangle^2
\nonumber\\
&&\hspace{2em}+15\langle\bar{s}s\rangle\langle\bar{u}u\rangle^2+\langle\bar{u}u\rangle
\langle\bar{s}s\rangle^2+\langle\bar{u}u\rangle\langle\bar{d}d\rangle^2
-8\langle\bar{d}d\rangle\langle\bar{s}s\rangle^2-8\langle\bar{s}s\rangle\langle\bar{d}d\rangle^2
\Big)\sp\ln{(-p^2)}
\nonumber\\
&&-\frac{\langle\alpha_s\pi^{-1}G^2\rangle}{3^3\!\times\!2^{12}\pi^2}
\Big(\langle\bar{u}u\rangle^2+\langle\bar{u}u\rangle\langle\bar{s}s\rangle\Big)
\sp\ln{(-p^2)}
\nonumber\\
&&-\frac{1}{3\!\times\!2^{17}\pi^4}\Big(
-11\langle\bar{u}g_s\vec{\sigma}\!\cdot\!\vec{G}u\rangle^2+207
\langle\bar{u}g_s\vec{\sigma}\!\cdot\!\vec{G}u\rangle
\langle\bar{d}g_s\vec{\sigma}\!\cdot\!\vec{G}d\rangle
\Big)\sp\ln{(-p^2)}
\nonumber\\
&&-\frac{1}{3^2\!\times\!2^9\pi^2}\Big(12\langle\bar{u}u\rangle^2
\langle\bar{u}g_s\vec{\sigma}\!\cdot\!\vec{G}u\rangle
-5\langle\bar{u}u\rangle^2\langle\bar{s}g_s\vec{\sigma}\!\cdot\!\vec{G}s\rangle
-5\langle\bar{u}u\rangle^2\langle\bar{d}g_s\vec{\sigma}\!\cdot\!\vec{G}d\rangle
\nonumber\\
&&\hspace{2em}+52\langle\bar{u}u\rangle\langle\bar{d}d\rangle
\langle\bar{s}g_s\vec{\sigma}\!\cdot\!\vec{G}s\rangle
+21\langle\bar{u}u\rangle\langle\bar{d}d\rangle\langle\bar{u}g_s\vec{\sigma}\!\cdot\!\vec{G}u\rangle
\nonumber \\
&&\hspace{10em}
+21\langle\bar{u}u\rangle\langle\bar{d}d\rangle\langle\bar{d}g_s\vec{\sigma}\!\cdot\!\vec{G}d\rangle
\Big)\ln{(-p^2)}
\nonumber\\
&&-\frac{\sp}{3^3\!\times\!2^4p^2}\Big(\langle\bar{u}u\rangle^2\langle\bar{s}s\rangle^2
+\langle\bar{u}u\rangle\langle\bar{s}s\rangle^3+\langle\bar{u}u\rangle\langle\bar{d}d\rangle^3
+\langle\bar{u}u\rangle\langle\bar{d}d\rangle\langle\bar{s}s\rangle^2\Big)
\nonumber\\
&&+\frac{m_u\sp}{3^3\!\times\!2^{11}\pi^2p^2}\Big(
6\langle\bar{u}u\rangle^2\langle\bar{u}g_s\vec{\sigma}\!\cdot\!\vec{G}u\rangle
+54\langle\bar{d}d\rangle^2\langle\bar{u}g_s\vec{\sigma}\!\cdot\!\vec{G}u\rangle
+54\langle\bar{s}s\rangle^2\langle\bar{u}g_s\vec{\sigma}\!\cdot\!\vec{G}u\rangle
\nonumber\\
&&\hspace{2em}+123\langle\bar{u}u\rangle^2\langle\bar{d}g_s\vec{\sigma}\!\cdot\!\vec{G}d\rangle
+123\langle\bar{u}u\rangle^2\langle\bar{s}g_s\vec{\sigma}\!\cdot\!\vec{G}s\rangle
-168\langle\bar{d}d\rangle^2\langle\bar{s}g_s\vec{\sigma}\!\cdot\!\vec{G}s\rangle
\nonumber\\
&&\hspace{2em}-168\langle\bar{s}s\rangle^2\langle\bar{d}g_s\vec{\sigma}\!\cdot\!\vec{G}d\rangle
+259\langle\bar{u}u\rangle\langle\bar{s}s\rangle
\langle\bar{u}g_s\vec{\sigma}\!\cdot\!\vec{G}u\rangle
+259\langle\bar{u}u\rangle\langle\bar{d}d\rangle
\langle\bar{u}g_s\vec{\sigma}\!\cdot\!\vec{G}u\rangle
\nonumber\\
&&\hspace{2em}-27\langle\bar{u}u\rangle\langle\bar{s}s\rangle\langle\bar{s}g_s\vec{\sigma}\!\cdot\!
\vec{G}s\rangle
-27\langle\bar{u}u\rangle\langle\bar{d}d\rangle\langle\bar{d}g_s\vec{\sigma}\!\cdot\!
\vec{G}d\rangle-90\langle\bar{u}u\rangle\langle\bar{d}d\rangle
\langle\bar{s}g_s\vec{\sigma}\!\cdot\!\vec{G}s\rangle
\nonumber\\
&&\hspace{2em}-90\langle\bar{u}u\rangle\langle\bar{s}s\rangle
\langle\bar{d}g_s\vec{\sigma}\!\cdot\!\vec{G}d\rangle
-500\langle\bar{d}d\rangle\langle\bar{s}s\rangle\langle\bar{u}g_s\vec{\sigma}\!\cdot\!\vec{G}u
\rangle-408\langle\bar{d}d\rangle\langle\bar{s}s\rangle\langle\bar{s}g_s\vec{\sigma}\!\cdot
\!\vec{G}s\rangle
\nonumber\\
&&\hspace{2em}-408\langle\bar{d}d\rangle\langle\bar{s}s\rangle\langle\bar{d}g_s\vec{\sigma}\!\cdot
\!\vec{G}d\rangle
\Big)+\{\textrm{flavor cyclic rotation}\}.
\label{aeq3}
\end{eqnarray}



\begin{thebibliography}{99}
\bibitem{Yao:2006px} W.~M.~Yao {\it et al.}  [Particle Data Group], J.\ Phys.\ G {\bf 33}, 1 (2006).
\bibitem{Reinders:1978mv} L.~J.~Reinders, J.\ Phys.\ G {\bf 4}, 1241 (1978).
\bibitem{Reinders:1980af} L.~J.~Reinders, {\it  In *Toronto 1980, Proceedings, Baryon 1980*, 203-207}
\bibitem{Sakurai:1960ju} J.~J.~Sakurai, Annals Phys.\  {\bf 11}, 1 (1960).
\bibitem{Dalitz:1967fp} R.~H.~Dalitz, T.~C.~Wong and G.~Rajasekaran, Phys.\ Rev.\  {\bf 153}, 1617 (1967).
\bibitem{YA1} T.~Yamazaki and Y.~Akaishi, Proc. Jpn. Acad. B83. 144 (2007).
\bibitem{Akaishi:2002bg} Y.~Akaishi and T.~Yamazaki, Phys.\ Rev.\  C {\bf 65}, 044005 (2002).
\bibitem{Yamazaki:2002uh} T.~Yamazaki and Y.~Akaishi, Phys.\ Lett.\  B {\bf 535}, 70 (2002).
\bibitem{Shifman:1978bx} M.~A.~Shifman, A.~I.~Vainshtein and V.~I.~Zakharov, Nucl.\ Phys.\  B {\bf 147}, 385 (1979).
\bibitem{Reinders:1984sr} L.~J.~Reinders, H.~Rubinstein and S.~Yazaki, Phys.\ Rept.\  {\bf 127}, 1 (1985).
\bibitem{Narison:1989aq} S.~Narison, World Sci.\ Lect.\ Notes Phys.\  {\bf 26}, 1 (1989).
\bibitem{Narison:2002pw} S.~Narison, Camb.\ Monogr.\ Part.\ Phys.\ Nucl.\ Phys.\ Cosmol.\  {\bf 17}, 1 (2002) [arXiv:hep-ph/0205006].
\bibitem{Liu:1984dp} J.~P.~Liu, Z.\ Phys.\  C {\bf 22}, 171 (1984).
\bibitem{Leinweber:1989hh} D.~B.~Leinweber, Annals Phys.\  {\bf 198}, 203 (1990).
\bibitem{LSR3} H.~Kim and S.~H.~Lee, Z.\ Phys.\ A {\bf 357}, 425 (1997).
\bibitem{Choe:1997wz} S.~Choe, Eur.\ Phys.\ J.\  A {\bf 3}, 65 (1998).
\bibitem{Jido:1996zw} D.~Jido and M.~Oka, arXiv:hep-ph/9611322.
\bibitem{Sugiyama:2007sg} J.~Sugiyama, T.~Nakamura, N.~Ishii, T.~Nishikawa and M.~Oka, Phys.\ Rev.\ D {\bf 76}, 114010 (2007).
\bibitem{DeGrand:1976kx} T.~A.~DeGrand and R.~L.~Jaffe, Annals Phys.\  {\bf 100}, 425 (1976).
\bibitem{Strottman:1979qu} D.~Strottman, Phys.\ Rev.\  D {\bf 20}, 748 (1979).
\bibitem{Jido:1996ia} D.~Jido, N.~Kodama and M.~Oka, Phys.\ Rev.\  D {\bf 54}, 4532 (1996).
\bibitem{Nemoto:2003ft} Y.~Nemoto, N.~Nakajima, H.~Matsufuru and H.~Suganuma, Phys.\ Rev.\  D {\bf 68}, 094505 (2003).
\bibitem{Melnitchouk:2002eg} W.~Melnitchouk {\it et al.}, Phys.\ Rev.\  D {\bf 67}, 114506 (2003).
\bibitem{Lee:2002gn} F.~X.~Lee, S.~J.~Dong, T.~Draper, I.~Horvath, K.~F.~Liu, N.~Mathur and J.~B.~Zhang, Nucl.\ Phys.\ Proc.\ Suppl.\  {\bf 119}, 296 (2003).
\bibitem{Burch:2006cc} T.~Burch, C.~Gattringer, L.~Y.~Glozman, C.~Hagen, D.~Hierl, C.~B.~Lang and A.~Schafer, Phys.\ Rev.\  D {\bf 74}, 014504 (2006).
\bibitem{Ishii:2007ym} N.~Ishii, T.~Doi, M.~Oka and H.~Suganuma, Prog.\ Theor.\ Phys.\ Suppl. {\bf 168}, 598 (2007).
\end{thebibliography}
\end{document}